\documentclass[conference]{IEEEtran}

  	\usepackage[pdftex]{graphicx}
  	\graphicspath{{../pdf/}{../jpeg/}}
	\DeclareGraphicsExtensions{.pdf,.jpeg,.png}

	\usepackage[cmex10]{amsmath}
	\usepackage{mathabx}
	\usepackage{algorithmic}
	\usepackage{array}
	\usepackage{mdwmath}
	\usepackage{mdwtab}
	\usepackage{eqparbox}
	\usepackage{url}
        \usepackage{float}
        \usepackage{hyperref}

\begin{document}

\title{\LARGE \textbf{From Sound to Setting: AI-Based Equalizer Parameter Prediction for Piano Tone Replication}
}
 \author{Song-Ze, Yu}

\maketitle

\begin{abstract}
This project presents an AI-based system for tone replication in music production, focusing on predicting EQ parameter settings directly from audio features. Unlike traditional audio-to-audio methods, our approach generates interpretable parameter values—such as EQ band gains—that musicians can further adjust in their workflow. Using a dataset of piano recordings with systematically varied EQ settings, we evaluate both regression and neural network models. Results show that our neural network model achieves highly accurate parameter predictions, with a mean squared error of 0.0216 on multi-band tasks. The proposed system enables practical, flexible, and automated tone matching for music producers, laying the foundation for future extensions to more complex audio effects.

\end{abstract}

\IEEEoverridecommandlockouts

\IEEEpeerreviewmaketitle


\section{Introduction}
In modern music production, replicating the tone of a reference sound is a crucial but often tedious part of the workflow. While much of the creativity lies in composition and arrangement, tasks such as matching EQ and other effects settings are mostly technical and rule-based. This study explores an AI-based approach for predicting EQ parameters directly from audio, aiming to automate tone replication while keeping the process interpretable and flexible for music producers.


\section{Background \& Related Work}

\subsection{Academic and Commercial Solutions}
Tone replication and timbre transfer have attracted growing attention in both academic research and commercial applications. One influential work is Google’s \textbf{DDSP Tone Transfer}\cite{engel2020ddsp}, which uses deep learning to perform audio-to-audio timbre transfer by synthesizing new sounds that match the target timbre. While this end-to-end approach enables creative transformations, it does not offer direct access to the underlying parameters—like synthesizer settings, EQ bands, or effects—that musicians rely on for precise tone shaping. As a musician, I see tone replication as just a starting point; creative work always requires the freedom to adjust these parameters further. Without such control, even the most accurate AI-generated tone would not be practical for real production needs.

On the commercial side, tools like \textbf{iZotope Ozone} offer automated EQ and mastering suggestions using proprietary algorithms. However, these systems are closed-source, and the underlying prediction mechanisms are not publicly disclosed, making it difficult for researchers or independent developers to build upon or evaluate them.

\subsection{Practical Tone Replication Workflows}
In practice, tone replication in music production can be approached in three main ways: 
\begin{enumerate}
    \item \textit{Directly sampling and reusing existing audio segments:} \\
    This approach is effective only when the goal is to reproduce the reference sound exactly, with no further adjustment. It offers no flexibility for creative modification or integration with new material.
    \item \textit{Transforming an existing sound to resemble a reference through EQ and effects}:\\
    This method, often described as a form of reverse engineering, involves adjusting parameters such as EQ, compressors, reverb, modulations to shape the frequency spectrum and dynamics of a sound, bringing it closer to the target tone.
    \item \textit{Synthesizing new sounds from scratch:}\\
    While theoretically powerful, synthesis-from-scratch with AI requires not only high technical complexity but also the ability to produce adjustable, interpretable parameters—something that end-to-end audio-to-audio neural models typically lack.
\end{enumerate}

\subsection{Our Approach}
Among these, using EQ and effects to approximate a target tone is both common and practical, as it allows producers to flexibly adapt and blend sounds within their workflow. Our work focuses on automating this reverse engineering process: by leveraging supervised learning, we predict the optimal EQ settings needed for piano tone replication, effectively treating the task as a form of inverse digital signal processing.


\section{Methods}
\subsection{Dataset Creation}
\noindent 1) \textbf{Piano Audio Database:}

To build a high-quality dataset for tone replication, I first recorded a series of MIDI performances covering every key from C0, G0, C1, G1, …, up to G7. These were rendered using the Logic Pro Steinway Grand Piano virtual instrument to generate .wav audio files. This approach allows for future extensibility—different instruments and playing styles can be added easily by replacing the MIDI instrument.

In addition to single notes, I also recorded useful musical elements such as C1–G6 scales, accented notes, and practical excerpts from pieces like Mozart’s K545, the Turkish March, and sections from pop songs and my own compositions. While these files have not been used in the current work, they serve as a foundation for future dataset expansion and available on GitHub.

\noindent2) \textbf{EQ Replication Data:}

For the main EQ regression experiments, I selected a core set of single-note audio files (C0–G7) to cover the full range of the piano. To ensure that the ground truth EQ parameters reflected actual music production practice, I wrote a custom ReaScript within \textbf{Reaper} to systematically apply EQ adjustments using the built-in EQ plugin.

Based on recommendations from \textit{The Mixing Engineer’s Handbook (2nd ed.)} \cite{oireilly2006mixing}, I chose five frequency bands that are both musically relevant and commonly adjusted by mixing engineers for piano. The low (80 Hz) and high (10,000 Hz) bands use shelf filters for practical reasons, while the mid bands use bell filters.

\begin{table}[ht]
\centering
\caption{EQ Bands Selected for Tone Replication Experiments}
\begin{tabular}{|c|c|c|l|}
\hline
\textbf{Band}    & \textbf{Frequency (Hz)} & \textbf{Filter Type} & \textbf{Mixing Purpose}   \\
\hline
EQ\_80    & 80     & Low-shelf   & Body, fullness           \\
EQ\_240   & 240    & Bell        & Warmth, presence         \\
EQ\_2500  & 2500   & Bell        & Attack, clarity          \\
EQ\_4000  & 4000   & Bell        & Presence, bite           \\
EQ\_10000 & 10000  & High-shelf  & Air, brightness          \\
\hline
\end{tabular}
\label{tab:eq-bands}
\end{table}

To systematically cover a wide range of practical EQ settings, I generated audio samples with EQ gain values ranging from \textbf{-12 dB} to \textbf{+12 dB} in 1 dB increments for each band. This resulted in two types of datasets:
\begin{itemize}
    \item \textit{\textbf{Single-band sweep:}} For each of the five bands, 25 different gain settings were applied individually, resulting in $25 \times 5 = 125$ unique audio files.
    \item \textit{\textbf{Multi-band combination:}} For more realistic and challenging scenarios, all five bands were adjusted simultaneously using seven representative gain steps (\{-12, -8, -4, 0, 4, 8, 12\} dB), resulting in $7^5 = 16,\!807$ unique audio files.
\end{itemize}

\subsection{Dataset Preprocessing}

Instead of directly using raw audio spectra, I opted for a feature extraction approach to increase the potential for dataset extensibility and generalization to new audio sets. Audio features were extracted using librosa package in Python. The extracted features include:
	
\begin{itemize}
    \item \textbf{Spectral centroid:} Represents the “center of mass” of the spectrum and correlates with the perceived brightness of the sound.
    \item \textbf{Spectral bandwidth:} Measures the spread of the spectrum around its centroid; it can reflect how “broad” or “focused” the frequency content is, and is especially sensitive to the use of shelf filters.
    \item \textbf{Spectral rolloff:} The frequency below which a certain percentage (usually 85--95\%) of the spectral energy is contained; it is often associated with the cutoff point of the spectrum.
    \item \textbf{MFCC means (coefficients 0 to 12):} Mel-frequency cepstral coefficients are a widely used representation of timbral features. Including 13 coefficients increases flexibility for future applications involving longer or more complex audio.
    \item \textbf{RMS energy:} The root mean square energy of the signal, capturing the overall loudness or intensity.
\end{itemize}

\subsection{Model System Design}

\begin{figure}[H]
\centering
\includegraphics[width=3.0in]{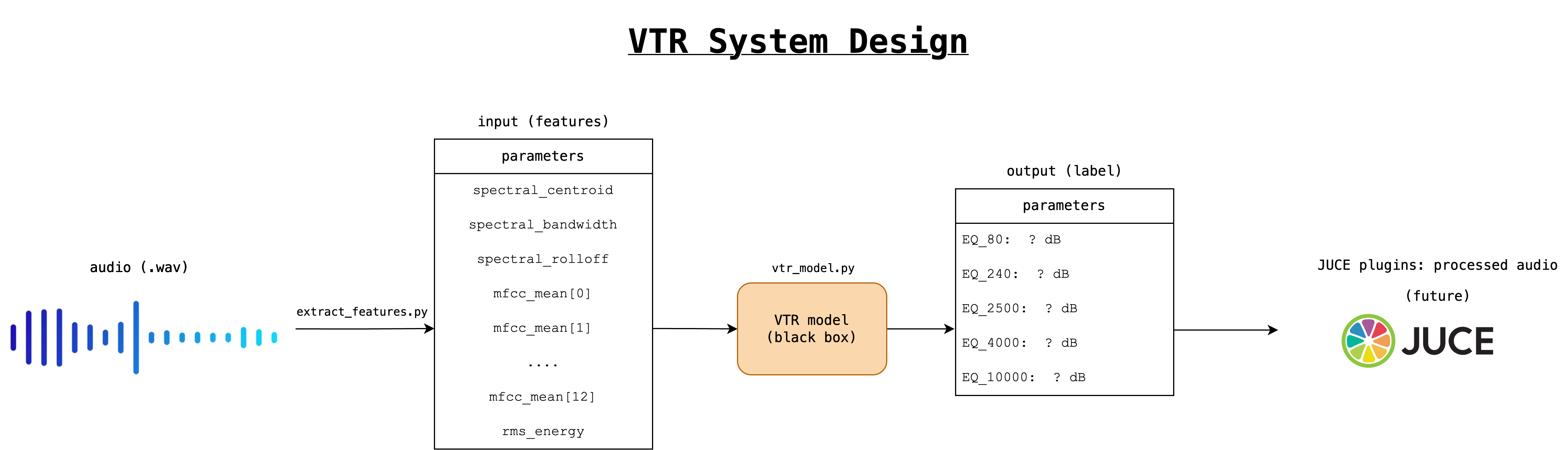} 
\caption{System Design of the VTR model.}
\label{fig:VTRModel}
\end{figure}

The overall system architecture of VTR (Vaclis Tone Replication) is illustrated in Figure~\ref{fig:VTRModel}. The workflow consists of the following modules:

\begin{enumerate}
    \item \textbf{Audio Input:} Raw audio files (.wav) are provided as input. These can be either single notes or musical phrases.
    \item \textbf{Feature Extraction:} Audio features relevant to timbre and spectral content---such as spectral centroid, spectral bandwidth, spectral rolloff, MFCC means (0 to 12), and RMS energy---are extracted using a custom Python script (\texttt{extract\_features.py}).
    \item \textbf{EQ Parameter Prediction Model:} The extracted features serve as input to the VTR model (\texttt{vtr\_model.py}), which predicts five target EQ parameters: EQ\_80, EQ\_240, EQ\_2500, EQ\_4000, and EQ\_10000 (in dB).
    \item \textbf{Plugin Application (Future work):} The predicted EQ parameters can be applied to audio processing plugins, such as those developed with JUCE, to replicate the target tone in real-world production workflows.
\end{enumerate}

\subsection{Prediction Model Architecture}

To map the extracted audio features to the five EQ parameters, we implemented two types of models:

\begin{itemize}
    \item \textbf{Linear Regression (Baseline):}
    As a baseline, we trained a standard linear regression model to directly predict EQ parameter values from the audio feature vector. This simple approach serves to verify the basic feasibility of parameter-level tone replication.
    
    \item \textbf{Feedforward Neural Network (FFNN):}
    Our main model is a feedforward neural network, implemented as a sequence of fully connected layers with ReLU activations. Specifically, the architecture is as follows:
    \begin{itemize}
        \item \texttt{nn.Linear(input\_dim, hidden\_dim)}
        \item \texttt{nn.ReLU()}
        \item \texttt{nn.Linear(hidden\_dim, hidden\_dim)}
        \item \texttt{nn.ReLU()}
        \item \texttt{nn.Linear(hidden\_dim, output\_dim)}
    \end{itemize}
    Here, \texttt{input\_dim} is the number of extracted audio features, \texttt{hidden\_dim} is the size of the hidden layer (tuned as a hyperparameter), and \texttt{output\_dim} is the number of target EQ bands (five).
\end{itemize}

Both models were trained using the mean squared error (MSE) loss function, and evaluated on a held-out test set.

\section{Results}
\subsection{Single-Band Sweep Results and Interpolation Test}

Since the single-band sweep dataset contains only 125 samples (5 bands $\times$ 25 gain values), and this setup does not reflect practical music production (as producers rarely adjust just one band at a time), we only conducted experiments with regression models and did not use neural networks.

\begin{figure}[H]
\centering
\includegraphics[width=2.9in]{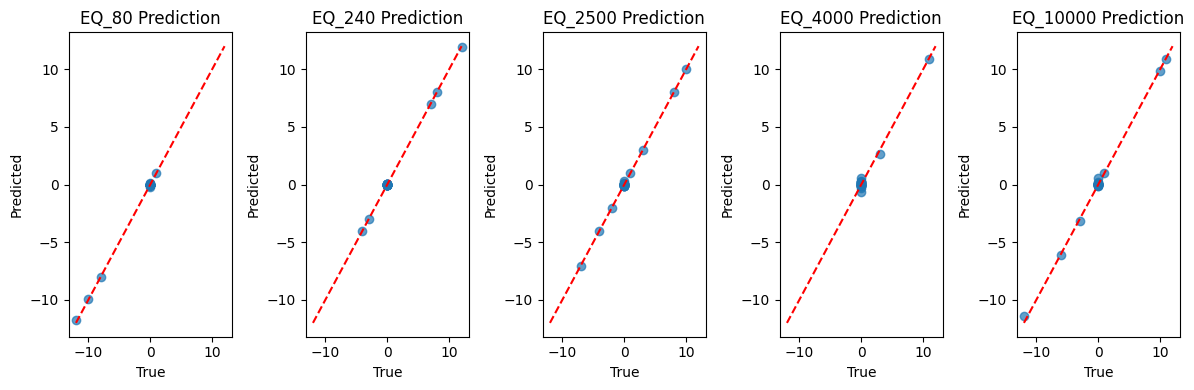} 
\caption{Single band regression result with resolution 1dB}
\label{fig:singleband-1db}
\end{figure}

We first tested the model with a fine 1~dB resolution across the full $-12$ to $+12$~dB range. The regression model achieved excellent results, with a mean squared error (MSE) of \textbf{0.0192} (see Figure~\ref{fig:singleband-1db}).

\begin{figure}[H]
\centering
\includegraphics[width=2.9in]{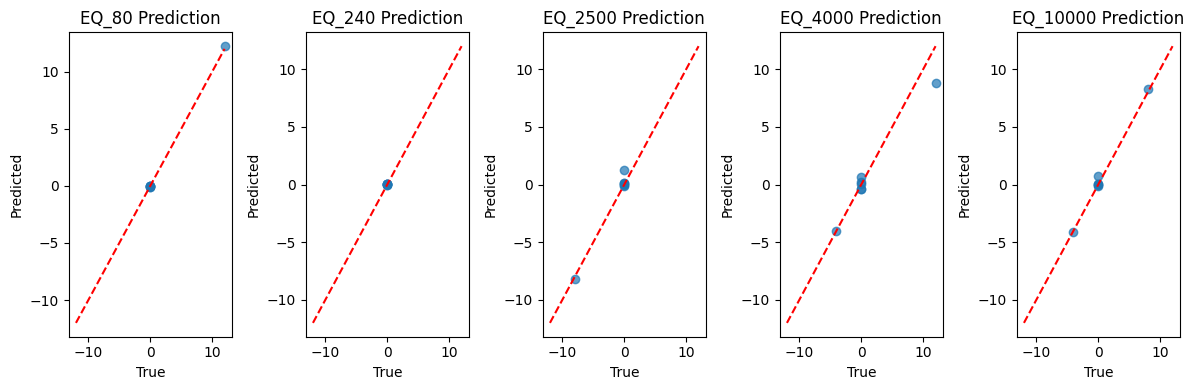} 
\caption{Single band regression result with resolution 4dB}
\label{fig:singleband-4db}
\end{figure}

Next, we tested a coarser 4~dB step size (using only $\{-12, -8, -4, 0, 4, 8, 12\}$~dB). As expected, the performance dropped significantly, with an MSE of \textbf{0.3745}(see Figure~\ref{fig:singleband-4db}). This is mainly due to the much smaller training set, which limits the model’s ability to capture the underlying relationships.

\begin{figure}[H]
\centering
\includegraphics[width=2.9in]{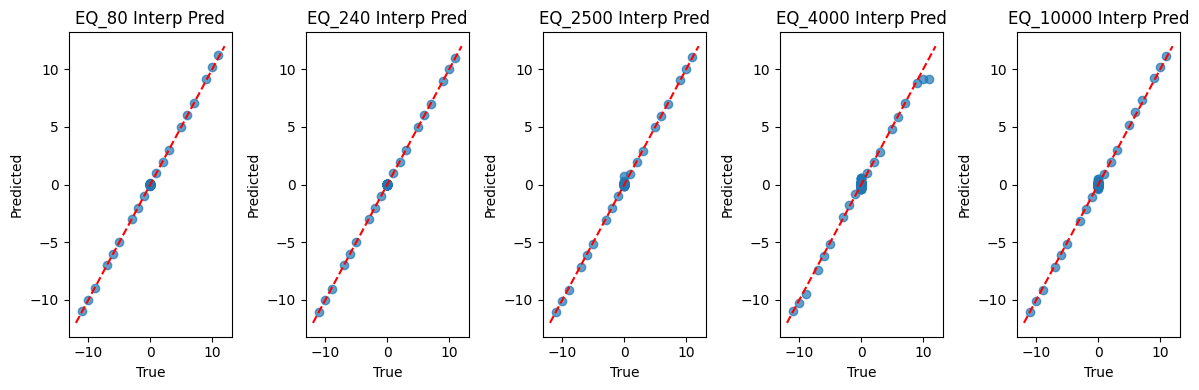} 
\caption{Interpolation result with resolution 4dB}
\label{fig:singleband-4db-test}
\end{figure}

To further evaluate the model’s interpolation ability, we trained on just the 4~dB increments and used the remaining in-between values as a validation set. The model still achieved a low interpolation MSE of \textbf{0.0272}(see Figure~\ref{fig:singleband-4db-test}), indicating it can generalize smoothly between trained points, even with sparser data. 

This result also shows that it is unnecessary to use excessively fine step sizes for practical training; the model is capable of learning smooth interpolation between the points. Therefore, in the subsequent multi-band dataset, we adopted a 4~dB step size for efficiency and practicality.

\subsection{Multi-Band Combination Results}

The multi-band dataset contains a total of $7^5 = 16,\!807$ samples, covering all possible combinations of gain settings at five frequency bands with 4~dB resolution ($\{-12, -8, -4, 0, 4, 8, 12\}$~dB). This setup better reflects practical mixing scenarios, where multiple EQ bands are adjusted simultaneously.
\begin{table}[h]
\centering
\caption{Mean Squared Error (MSE) Comparison on Multi-band Dataset}
\begin{tabular}{|l|c|}
\hline
\textbf{Model}                  & \textbf{Mean Squared Error (MSE)} \\
\hline
Linear Regression (baseline)    & 0.5354 \\
Random Forest Regressor         & 1.2740 \\
Fully Connected Neural Network  & \textbf{0.0216} \\
\hline
\end{tabular}
\label{tab:mse-results}
\end{table}

\begin{figure}[H]
\centering
\includegraphics[width=3.0in]{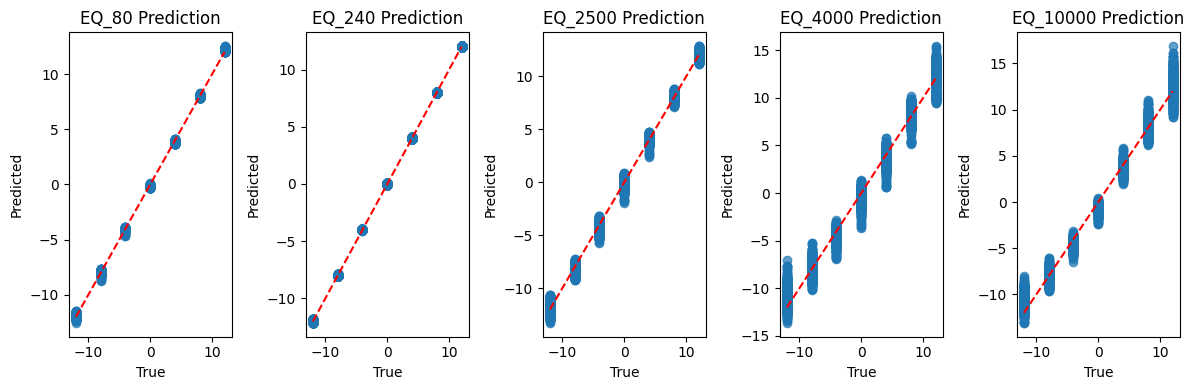} 
\caption{Regression result}
\label{fig:multiband-regression}
\end{figure}

We first trained a linear regression model as a baseline. Although the model could generally follow the trend, the performance was limited, achieving a mean squared error (MSE) of \textbf{0.5354} (see Figure~\ref{fig:multiband-regression}). 

\begin{figure}[H]
\centering
\includegraphics[width=3.0in]{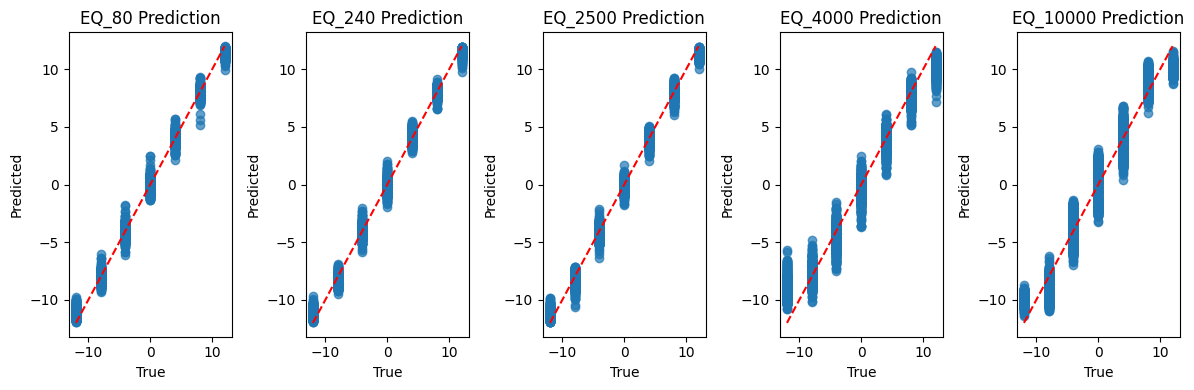} 
\caption{Random forest regressor result}
\label{fig:multiband-regression-rf}
\end{figure}

We also experimented with a random forest regressor, but its MSE was even higher (\textbf{1.2740}) (see Figure~\ref{fig:multiband-regression-rf}), likely due to overfitting and limited generalization in high-dimensional parameter spaces.

\begin{figure}[H]
\centering
\includegraphics[width=3.0in]{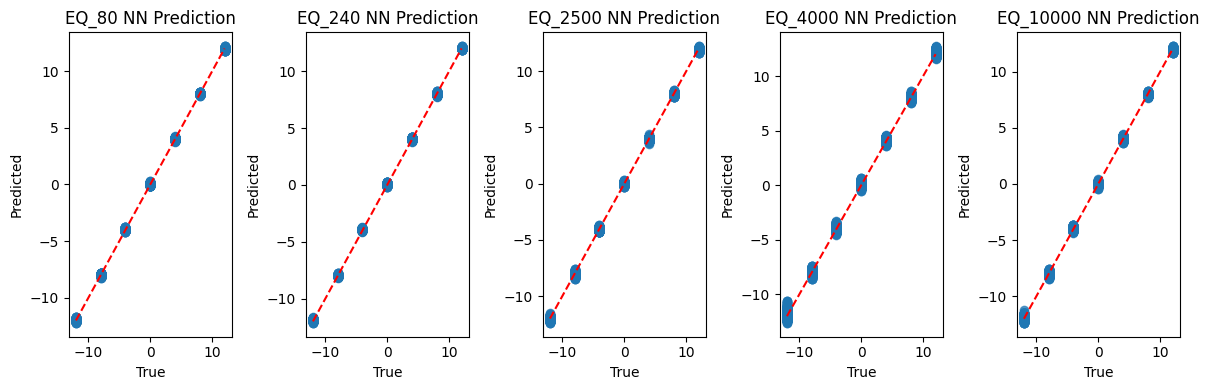} 
\caption{Feedforward Neural Network result}
\label{fig:multiband-nn}
\end{figure}

To further improve performance, we implemented a fully connected neural network (NN) with two hidden layers and ReLU activation. The NN model significantly outperformed traditional regressors, achieving a best MSE of \textbf{0.0216} (see Figure~\ref{fig:multiband-nn}). The predicted EQ values closely matched the ground truth, even under complex multi-band conditions.


\section{Discussion}

\subsection{VTR Model Result Analysis}
Our results demonstrate that \textbf{supervised learning models can effectively predict EQ settings from audio features}. In the single-band scenario, regression models perform well even with a very limited dataset, achieving near-perfect accuracy. However, in the more complex multi-band setting, neural networks are essential for handling the increased dimensionality and nonlinearity, and significantly outperform traditional regression methods.

\subsection{Limitations}
\begin{itemize}
    \item \textit{Scope of Tone Replication:} \\Tone replication is an inherently complex task. EQ alone---even with five carefully chosen bands---\textbf{cannot fully capture the nuances of a target sound}. Our current focus on EQ serves as a foundational step, but comprehensive tone matching will require extending the system to dynamic effects such as compressors, reverbs, and modulations. This represents a major challenge and will be the focus of future work.
    \item \textit{Evaluation Methodology:} \\Our evaluation thus far relies solely on quantitative metrics such as MSE and scatter plots comparing predicted and ground truth EQ values. While these metrics are informative, they do not fully capture perceptual similarity. Future validation should include larger, \textbf{more diverse datasets} and \textbf{listening tests} to assess perceived tone replication quality.
\end{itemize}

\subsection{Future Work}
\begin{itemize}
    \item Short-term: \\Integrate the trained prediction models into a usable tool---either as a ReaScript script or a JUCE-based \textbf{VST plugin}. The ultimate goal is to create an AI-driven VTR (Vaclis Tone Replication) plugin that can be applied in practical music production workflows.\\
    \item Medium-term: \\Extend the system to handle additional audio processing parameters, beginning with \textbf{compressors and modulation} effects, and investigate models capable of simultaneously predicting multiple effect types.\\
    \item Long-term: \\Develop a unified system for comprehensive tone replication, capable of \textbf{predicting and controlling all key parameters} involved in sound design. All datasets and scripts will be released as open source to support future research and community-driven improvements.
\end{itemize}


\section{Conclusion}

We have presented a practical system for parameter-level tone replication, using supervised learning to predict EQ settings from audio features. Our experiments show that this approach is both accurate and efficient, especially when using neural networks for multi-band scenarios. The results confirm that feature-based modeling is a viable alternative to black-box audio-to-audio methods, offering interpretable and flexible tools for real-world music production. Future work will focus on expanding to more effect types and developing a user-ready plugin for practical use.



\section*{Data and Code Availability}
The code for dataset downloading, preprocessing, model training, and result presentation is available in our GitHub repository: \url{https://github.com/vaclisinc/Vaclis_Tone_Replication}.

Please refer to the README file in the repository for detailed instructions on how to run the code and reproduce the results.


\nocite{engel2020ddsp}
\nocite{oireilly2006mixing}
\nocite{hollemans2023synth}
\nocite{hollemans2024complete}
\nocite{yeeking2024ai}
\bibliographystyle{IEEEtran}
\bibliography{IEEEabrv,biblio_traps_dynamics}

\end{document}